\documentclass[10pt,a4paper]{article}
\usepackage{graphicx}
\usepackage{amssymb}
\usepackage{amsmath}
\usepackage{multirow}

\newtheorem{theo}{Theorem}

\begin{document}
\begin{flushleft}
{\Large
\textbf{A Max-Plus Model of Ribosome Dynamics During mRNA Translation}
}

\vspace{15pt}

Chris A. Brackley$^{1}$, David Broomhead$^2$ M. Carmen Romano$^{1,3}$, and Marco Thiel$^{1}$.

\vspace{5pt}
\begin{footnotesize}
1 \textit{Institute for Complex Systems and Mathematical Biology, SUPA, University of Aberdeen, Aberdeen, AB24 3UE, UK}
\\
2 \textit{School of Mathematics, Alan Turing Building, The University of Manchester, Oxford Road, Manchester M13 9PL, UK}
\\
3 \textit{Institute of Medical Sciences, Foresterhill, University of Aberdeen, Aberdeen, AB25 2ZD, UK}
\\
\end{footnotesize}
\end{flushleft}

\begin{abstract}
We examine the dynamics of the translation stage of cellular protein production, in which ribosomes move uni-directionally along mRNA strands building an amino acid chain as they go. We describe the system using a timed event graph - a class of Petri net useful for studying discrete events which take a finite time. We use max-plus algebra to describe a deterministic version of the model, calculating the protein production rate and density of ribosomes on the mRNA. We find exact agreement between these analytical results and numerical simulations of the deterministic case.
\end{abstract}


\section{Introduction}

Messenger RNA (mRNA) translation is one of the steps in protein
production in cells \cite{Albertsbook}. mRNAs are single strands of nucleotides which are
transcribed from the DNA. The sequence of nucleotides, grouped in triplets called codons, holds the code for a specific chain of amino acids that makes up a protein. Translation is performed by molecular machines
called ribosomes, which scan
along the mRNA adding amino acids to a growing chain which will become the
protein. The rates at which different proteins are
produced are crucial in determining how a cell grows and
functions. 

Various statistical models have been used to describe and understand the translation process. In this paper we propose a new model which can be analysed rather completely using algebra on the max-plus semi-ring. The most convenient way to present the model is to write it as a deterministic Petri net \cite{Murata1989}. This can then be analysed completely as a linear dynamical system when written in terms of the max-plus algebra \cite{MPbook}. The model gives a more realistic representation of the movement of ribosomes than that of many previous studies, and explicitly considers the exact genetic coding sequence as well as the fact that different codons are translated at different rates. We perform numerical simulations of the Petri net, and find the results to exactly match the behaviour predicted by the max-plus algebra.

Although the general sequence of events in translation is understood,
there are many questions which remain unanswered. Translation begins
when a ribosome binds to the end of an mRNA strand - a process known as initiation. In this paper we focus on the second stage of translation, which is known as elongation. Here the ribosome moves uni-directionally along the mRNA, pausing at each codon to recruit an amino acid, which is then added to the growing chain. Amino acids are transported within the cytoplasm by carrier molecules called transfer RNAs (tRNAs). Different species of tRNA carry different amino acids, and different codons correspond to different tRNAs. The different tRNAs appear in different concentrations, so the time which a ribosome waits for the required tRNA to arrive differs from codon to codon \cite{Robinson1984,Sorensen1989}. Since it is often the case that several ribosomes are bound to the mRNA at the same time, the distribution of waiting times can result in traffic jams if the progress of one ribosome is obstructed by another. This picture is complicated by the fact that whilst there are only 20 common amino acids, in the model organism \emph{Saccharomyces Cerevisiae} there are 41 species of tRNA, i.e., there is redundancy in the genetic code. For some amino acids there are multiple tRNAs; furthermore it is often that case that there is a highly abundant tRNA and a rare tRNA which code for the same amino acid. Sometimes a ``slow codon'', which will cause a long pause in elongation, is used when a fast codon corresponding to the same amino acid is available. The interest for biologists is therefore to understand how slow codons effect protein production rates and the use of ribosomes \cite{Brockmann2007}. Translation reaches its completion in the termination stage where, through the binding of release factors, the ribosome disassociates from the mRNA and the amino acid chain is released ready for folding or further processing. The quantities of interest which we shall take from the model are the time interval between successive amino acid chains being completed, which we will call the protein production time, and the ribosome occupation density of the mRNA. 

In the next section we introduce Petri nets, and detail how these can be used to describe translation. Then in Sec. \ref{sec:maxplus} we introduce the max plus algebra, giving a brief survey of the salient facts required in the rest of the paper. In Sec. \ref{sec:analytic} we detail the max-plus treatment of the Petri net describing translation. These results are then compared to numerical simulations: we first consider simple ``designer mRNAs'' where each codon corresponds to the same tRNA, before treating realistic sequences taken from the \emph{S. Cerevisiae} genome. Finally we consider how introducing stochasticity into the model is likely to alter the results, and how the present work relates to previous models of translation (specifically the totally asymmetric simple exclusion process or TASEP).

\section{Timed Petri Nets}\label{sec:petri}

Petri nets (attributed to C. A. Petri \cite{petrithesis}) are a scheme where a sequence of discrete events is described on a
network. We give a brief introduction here, but for a detailed description refer the reader to
Ref. \cite{Murata1989} and references therein.

\begin{figure}
\centering
\includegraphics{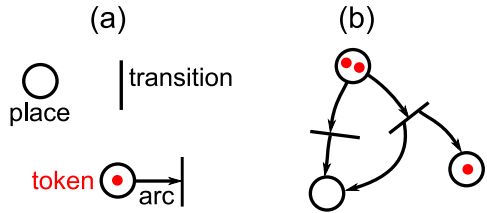}
\caption{Colour on-line. (a) Petri net components: places, transitions, arcs and
  tokens. (b) A simple Petri net. \label{petri_intro_diag} }
\end{figure}

A Petri net is a directed graph which contains two types of nodes:
places and transitions. These are shown diagrammatically in
Fig. \ref{petri_intro_diag}(a). Directed arcs connect places with
transitions, but not places with places or transitions
with transitions. Places can contain objects called tokens. A
transition is said to be active when each of the upstream places it is
connected to contains at least one token. Figure \ref{petri_intro_diag}(b) shows an
example Petri net. When a transition is active
it can fire; upon firing one token is removed from all places upstream
of the transition, and one token is added to all places downstream of
the transition (clearly there is no implicit conservation of tokens since the number of upstream places need not equal the number of downstream places). Events unfold in a discrete manner. Petri nets are often
used to model systems where events occur (transitions fire) given that
a set of conditions are fulfilled (tokens are present).

A timed Petri net is an extension to this framework in which a waiting time is
attached to each place. Whenever a token is put into a place, it will
only contribute to the activation of transitions once a time $\tau$
associated with that place has elapsed. This can be thought of as a
timer on the token which is started as it enters a place. As soon as
a transition becomes active it fires. In this way
the Petri net describes discrete events occurring in continuous time as
conditions are fulfilled. Unlike some other
models (for example Monte Carlo based simulations) events can occur
simultaneously.

\begin{figure}
\centering
\includegraphics[width=15cm]{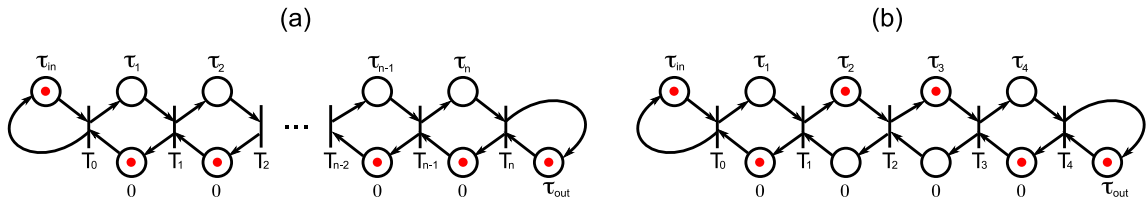}
\caption{Colour on-line. (a) A Petri net describing mRNA translation. Each pair (upper and
  lower) of places represent one codon. Tokens (red) in the bottom
  row represent vacant codons; in the top row they represent
  ribosomes reading that codon. The initial condition with no
  ribosomes on the mRNA is shown. (b) An example Petri net for a short mRNA of length $n=4$ codon, with a token configuration representing ribosomes occupying codons 2 and 3. \label{mRNA_petri} }
\end{figure}

We can describe the movement of ribosomes along the mRNA as a timed
Petri net, and this is shown in Fig. \ref{mRNA_petri}(a). A sequence of codons of length $n$ is
represented by two rows of places, with each pair (upper and lower)
representing a codon. A token in the top row indicates that a
ribosome is decoding that codon, and a token in the bottom row
indicates a vacant codon. Having this two row structure ensures that
there can only be one token in each place; i.e., one ribosome
translating each codon. This of course requires that the initial
state also satisfies this condition. A suitable initial state
is having all top row places empty and all bottom row places
containing one token; this corresponds to a mRNA initially free of
ribosomes.

 The firing of a transition $\mathrm{T}_i$ represents the movement of a ribosome from codon $i$ to codon $i+1$. There are $n+1$ transitions, labelled $i=0,\dots n$, with the firing of the zeroth and $n$th transitions representing initiation and termination events respectively. At the leftmost end of the Petri net we have only an upper row place;
this represents a continuous supply of ribosomes ready to begin
translation. At the rightmost end of the Petri net there is only a
lower row place, representing the cytoplasm which is always ready to
accept ribosomes. These places are kept full with a token via an arc
to and from the same transition.

The time associated with each upper row place corresponds to the time
it takes a ribosome to capture a tRNA of the species corresponding to
that codon. Generally the arrival times of tRNAs can be represented by a set of independent Poisson processes, however a deterministic version of the model can be obtained by replacing these random times with their means. In the present paper we concentrate on the deterministic version; the consequences of introducing stochasticity are discussed in Sec. \ref{sec:stoch}. We assume that the mean codon waiting time is proportional to the concentration of the species of tRNA corresponding to that codon. We denote the waiting
time for the $i$th codon counting from the left
$\tau_i$. We assume that when a ribosome moves forward the codon it was covering is immediately vacant, and the time it takes for the ribosome to physically move is negligible; thus we associate
a time zero with each of the lower row of places. In reality a ribosome
actually covers more than just the codon it is translating, but for
simplicity we do not take this into account here. The first place in the chain - corresponding to
initiation - has a time $\tau_{\mathrm{in}}$ associated with
it. Biologically this corresponds to the average time interval between attempts for a
ribosome to begin translation, and will depend on the availability of ribosomes and various initiation factors, as well as the presence of any secondary structures in the non-coding leader region of the mRNA. Since our focus here is on elongation, we assume that $\tau_{\mathrm{in}}$ is constant in time, treating it as a control parameter. The final place in the chain has an associated time $\tau_{\mathrm{out}}$ which is the time a ribosome waits at the end of the mRNA before releasing the completed protein and detaching from the mRNA. Biologically this could be linked to the availability of a number of release factors, and again we treat it as a control parameter. The model is therefore parametrised by the initiation and termination waiting times $\tau_{\mathrm{in}}$ and $\tau_{\mathrm{out}}$, and the set of internal tRNA capture waiting times $\{ \tau_i | 1\leq i \leq n\}$.

Figure \ref{mRNA_petri}(b) shows and example Petri net for a very (unrealistically) short mRNA of length $n=4$ codons, with ribosomes occupying codons 2 and 3. Transition $\mathrm{T}_2$ cannot fire because there is no token in its input place in the lower row (representing a vacancy). Transition $\mathrm{T}_3$ will fire as soon as the token in its upper row input place has been there for a time $\tau_3$. If the token in the place upstream of $\mathrm{T}_2$ (upper row place for codon 2) has been there for a time longer than $\tau_2$ when $\mathrm{T}_3$ fires, then $\mathrm{T}_2$ will also fire at this time. Thus if there is a slow codon (large $\tau_i$), when the corresponding transition fires, it could result in several other transitions simultaneously firing. 

\section{The max-plus semi-ring}\label{sec:maxplus}

If each place in a Petri net has exactly one upstream and one downstream
transition (as is the case in those depicted in Fig. \ref{mRNA_petri}), then
its dynamics can be described using max-plus algebra. Max-plus is an
algebra over the semi-ring $\mathbb{R}_{\mathrm{max}} \in \mathbb{R}
\cup -\infty$. In this section we give a brief overview of the theorems required in the rest of the paper, and refer the
reader to Refs. \cite{MPbook,Cohen1999} for further details and proofs.

In max-plus algebra the addition and multiplication operators
$\oplus$ and $\otimes$ are defined as
\begin{equation*}
a\oplus b=\max(a,b) ~~\mathrm{and}~~ a\otimes b=a+b,
\end{equation*}
where $a,b\in\mathbb{R}_{\mathrm{max}}$, and we define
$\epsilon=-\infty$ and $e=0$ which behave as zero and unity
respectively. Max-plus algebra refers to the set $\mathcal{R}_{\mathrm{max}}=(\mathbb{R}_{\mathrm{max}},\otimes,\oplus,\epsilon,e)$. The elements of $\mathcal{R}_{\mathrm{max}}$ have much the same properties
as in conventional algebra; for example
\begin{align*}
a\otimes b&=b\otimes a, \\
a \otimes (b\oplus c)&=a\otimes b \oplus a\otimes c,\\
a\otimes \epsilon&=\epsilon,\\
a\otimes e&=a,~\mathrm{etc.}
\end{align*}
We define Max plus powers in the natural way
\begin{equation*}
a^{\otimes x}=a\otimes a\otimes \dots \otimes a ~~x\mathrm{ ~times},
\end{equation*}
and note that in conventional algebra this corresponds to $x\times a$. We note that the $\otimes$ operator has an inverse which can be expressed as a negative power
\begin{equation*}
a \otimes b^{\otimes (-1)} = a-b,
\end{equation*}
but there is no inverse of the $\oplus$ operator.
We also use the following notation for sums and products over indices:
\begin{equation*}
\bigoplus_{i=1}^n a_i=\max\{a_i|1\leq i \leq n\} ~~\mathrm{and}~~
\bigotimes_{i=1}^n a_i=\sum_{i=1}^n a_i.
\end{equation*}

Vectors and matrices are also defined, and we denote by $[A]_{ij}$ the $i\mbox{-} j$th component of the matrix $A$. The sum and product of matrices
$A\in\mathbb{R}_{\mathrm{max}}^{n\times l}$ and $B\in\mathbb{R}_{\mathrm{max}}^{l\times m}$ are
then defined as
\begin{align*}
[A\oplus B]_{ij}&=a_{ij}\oplus b_{ij}, \\
[A\otimes B]_{ij}&=\bigoplus_{k=1}^l a_{ik}\otimes b_{kj},
\end{align*}
where $a_{ij}=[A]_{ij}$ and $b_{ij}=[B]_{ij}$.

Matrices can have associated eigenvectors and eigenvalues; i.e. a vector $\mathbf{u}$ satisfying 
\begin{equation*}
A\otimes\mathbf{u}=\lambda\otimes \mathbf{u},
\end{equation*}
is an eigenvector of $A$ associated with eigenvalue $\lambda$. Note that eigenvectors are not unique, i.e. if $\mathbf{u}$ is an eigenvector then so is $\alpha\otimes\mathbf{u}$ with $\alpha$ an arbitrary finite number. In general a matrix can have multiple associated eigenvalues.

The properties of a max-plus matrix can often be determined by considering the associated directed weighted graph. The graph associated with matrix $A$, denoted $\mathcal{G}(A)$, consists of a set of nodes and directed weighted arcs, where if there is a matrix element $[A]_{ij}$ with value $a_{ij}\neq\epsilon$, then there is an arc from node $j\rightarrow i$ with weight $a_{ij}$ (note that direction of the arc differs to that in conventional algebra and graph theory). A series of of one or more arcs between two nodes $i$ and $j$ is called a path from $i\rightarrow j$, and if there is a path $i\rightarrow i$ this is called a circuit and is denoted $\gamma$. A graph is said to be strongly connected if for any two different nodes there is a path between them, and a matrix $A$ is said to be irreducible if the corresponding graph $\mathcal{G}(A)$ is strongly connected. The circuit weight $w_\gamma$ of a circuit $\gamma$ is defined as the sum of the weights of all arcs in that circuit, and the circuit length $l_\gamma$ as the number of arcs in the circuit. The mean circuit weight is defined $\bar{w}_\gamma=w_\gamma/l_\gamma$.  If the maximum mean circuit weight in a graph is $\lambda$, then a circuit with a mean circuit weight equal to $\lambda$ is called a critical circuit. The critical graph corresponding to matrix $A$, denoted $\mathcal{G}^c(A)$, is defined as the sub-graph of $\mathcal{G}(A)$ containing only the nodes and arcs which are in the critical circuits. The cyclicity of a graph $\sigma_{\mathcal{G}}$ is defined as the greatest common divisor of the lengths of all of the circuits in that graph, and the cyclicity of a matrix $A$, $\sigma_A$ is equal to the cyclicity of the \emph{critical graph} of $A$, $\mathcal{G}^c(A)$. 
\begin{theo}\label{T:eigen}
Any irreducible matrix $A\in\mathbb{R}_{\mathrm{max}}^{n\times n}$ possesses one and only one eigenvalue $\lambda$, which is a finite number and is equal to the maximal mean circuit weight of circuits in the graph $\mathcal{G}(A)$. (See, for example, Theorem 2.9 in \cite{MPbook} for proof.) 
\end{theo}
Thus the eigenvalue of an irreducible matrix can be found by considering the corresponding graph. As noted above there is a whole continuum of eigenvectors associated with the eigenvalue of a max-plus matrix. Two eigenvectors $\mathbf{x},\mathbf{y}$ are co-linear if there exists a scalar $\alpha \in \mathbb{R}_{\mathrm{max}}$ such that $\mathbf{y}=\alpha \otimes \mathbf{x}$, and such vectors can be projected onto the same object in a projective space \cite{MPbook}. A given eigenvalue of the matrix $A$ can be associated with more that one linearly independent eigenvector, and the number of such eigenvectors can be found by considering the critical graph $\mathcal{G}^c(A)$.
\begin{theo}\label{mscs}
If the critical graph $\mathcal{G}^c(A)$ of an irreducible matrix $A$ has $k$ maximal strongly connected sub-graphs (m.s.c.s.), then $A$ has $k$ linearly independent eigenvectors. (For proof see, for example, Theorem 4 in \cite{Cohen1985}.)
\end{theo}

Linear equations such as 
\begin{equation}
\mathbf{x}=(A\otimes \mathbf{x})\oplus \mathbf{b}, \label{linear}
\end{equation}
where $A\in\mathbb{R}_{\mathrm{max}}^{n\times n}$ and $\mathbf{x},\mathbf{b}\in\mathbb{R}_{\mathrm{max}}^{n}$, can often be solved using an object called the Kleene star, defined as
\begin{equation}
A^*=\bigoplus_{k=0}^{\infty} A^{\otimes k}.
\end{equation}
The existence of $A^*$ can be proven if the graph $\mathcal{G}(A)$ has only non-positive circuit weights \cite{MPbook}. 
\begin{theo}\label{T:lin}
If $\mathcal{G}(A)$ has maximal mean circuit weight less than or equal to $e=0$, then $\mathbf{x}=A^*\otimes\mathbf{b}$, is a unique solution to Eq. (\ref{linear}). (For proof see, for example, Theorem 2.10 in \cite{MPbook}.)
\end{theo}

We shall see below that the problem studied in this paper involves sequences of vectors $\mathbf{x}(k), k\in\mathbb{N}$ described by a recurrence equation
\begin{equation*}
\mathbf{x}(k)=A \otimes\mathbf{x}(k-1),
\end{equation*}
or equivalently
\begin{equation}
\mathbf{x}(k)=A^{\otimes k} \otimes\mathbf{x}(0), \label{series0}
\end{equation}
for $k\geq0$ where $A\in\mathbb{R}_{\mathrm{max}}^{n\times n}$ is irreducible, and $\mathbf{x}(0)$ is some initialisation vector. 
\begin{theo}\label{T:cyc}
The cyclicity theorem states that there is an integer $K_0$ such that 
\begin{equation*}
A^{\otimes (k+\sigma_A)} = \lambda^{\otimes \sigma_A} \otimes A^{\otimes k}, ~~~k\geq K_0,
\end{equation*}
where $\lambda$ and $\sigma_A$ are the eigenvalue and cyclicity of the matrix $A$ respectively. (For proof see \cite{Cohen1985} or Theorem 3.9 in \cite{MPbook}.)
\end{theo}
This implies some periodicity in the powers of $A$; the periodic behaviour is characterised by $\lambda$ and $\sigma_A$, and $K_0$ is known as the transient time of $A$. By using this in Eq. (\ref{series0}) we find
\begin{align*}
x(k+\sigma) &= A^{\otimes (k+\sigma)} \otimes x(0)  \\
&= \lambda^{\otimes \sigma} \otimes A^{\otimes k} \otimes x(0)  \\
&= \lambda^{\otimes \sigma} \otimes x(k).
\end{align*}
If matrix $A$ has cyclicity $\sigma_A=1$, then we have
\begin{align}
x(k+1) &= A\otimes x(k) \nonumber \\
&= \lambda \otimes x(k), ~~~k\geq K_0,
\end{align}
i.e., $\mathbf{x}(k)$ (or equivalently $A^{\otimes k}\otimes \mathbf{x}(0)$ for any $\mathbf{x}(0)\in\mathbb{R}_{\mathrm{max}}^{n}$) is an eigenvector of $A$ for $k\geq K_0$. The effect of the initial condition has died out.

\section{Analysis of translation with the max-plus algebra}\label{sec:analytic}

A Petri net can be described using max plus algebra by defining the
matrices $A_0,A_1,\dots,A_M$ which depend on the initial conditions such
that $[A_m]_{ij}$ is equal to the maximum of the holding times
associated with the places between transitions $j$ and $i$, which
initially contain $m$ tokens. The dynamics are then described by the
vector $\mathbf{x}(k)$, where the $i$th component of the vector
is equal to the time at which the $i$th transition fires
for the $k$th time. This vector satisfies the recursion
equation
\begin{equation*}
\mathbf{x}(k)=A_0 \otimes \mathbf{x}(k) \oplus A_1\otimes
\mathbf{x}(k-1) \oplus \dots \oplus A_M\otimes \mathbf{x}(k-M).
\end{equation*}
For further details see \cite{MPbook,Baccelli1992,Gaubert1997}. 

In the case of the Petri net which describes mRNA translation (shown in Fig \ref{mRNA_petri}(a)), the structure and initial conditions are such that each place can contain either 0 or 1 tokens, i.e., in this case $M=1$ and the above equation reduces to
\begin{equation}
\mathbf{x}(k)=A_0 \otimes \mathbf{x}(k) \oplus A_1\otimes\mathbf{x}(k-1). \label{recurs}
\end{equation}
In the Petri net we have $n+1$ transitions labelled $0,\dots n$, so $\mathbf{x}(k)$ has components $\{x_i|0\leq i\leq n\}$. Likewise $A_0$ and $A_1$ are $(n+1)\times (n+1)$ matrices, with elements labelled by indices running from $0$ to $1$.
These matrices are 
\begin{equation*}
A_0= \left( \begin{array}{ccccc}
\epsilon & \epsilon & \epsilon & \cdots & \epsilon \\
\tau_1 & \epsilon & \epsilon & \cdots &\epsilon \\
\epsilon & \tau_2 & \epsilon & \cdots & \epsilon\\
\vdots &\ddots &\ddots &\ddots & \vdots\\
\epsilon & \cdots &\epsilon & \tau_n & \epsilon
\end{array} \right)
~~~\mathrm{and}~~~
A_1= \left( \begin{array}{ccccc}
\tau_{\mathrm{in}} & 0 & \epsilon & \cdots & \epsilon \\
\epsilon & \epsilon & 0 & \cdots &\epsilon \\
\epsilon & \epsilon& \ddots & \ddots & \epsilon\\
\vdots &\ddots &\ddots &\epsilon & 0\\
\epsilon & \cdots &\epsilon & \epsilon & \tau_{\mathrm{out}}
\end{array} \right).
\end{equation*}
Written in component form Eq. (\ref{recurs}) gives
\begin{align*}
x_0(k)&=\max \{ \tau_{\mathrm{in}}+x_0(k-1) , x_1(k-1) \},\\
x_i(k)&=\max \{ \tau_{i} + x_{i-1}(k), x_{i+1}(k-1) \} ~~\mathrm{for}~1\leq i< n,\\
x_n(k)&=\max \{ \tau_{n} + x_{n-1}(k), \tau_{\mathrm{out}} + x_n(k-1) \}.
\end{align*}
If we think of the integer $k$ as labelling the ribosomes ($x_i(k)$ gives the time at which the $k$th ribosome leaves the $i$th codon) then these equations make sense conceptually since a ribosome will leave codon $i$ either a time $\tau_{i}$ after it left codon $i-1$, or at the time when the $(k-1)$th ribosome leaves codon $i$ (i.e., time $x_{i+1}(k-1)$), whichever is latest.

Although the solution of the recursion equation (\ref{recurs}) is not straightforward, since there is no inverse to the $\oplus$ operator, by identifying the second term as a vector $\mathbf{b}=A_1\otimes\mathbf{x}(k-1)$, we note that this equation is of the form of Eq. (\ref{linear}), and so via Theorem \ref{T:lin} the solution is
\begin{equation*}
\mathbf{x}(k)=A_0^* \otimes A_1\otimes\mathbf{x}(k-1).
\end{equation*}
From that theorem we know that $A_0^*$ exists since the graph $\mathcal{G}(A_0)$ contains no circuits of positive weight; this can also be easily demonstrated by considering the first few powers of $A_0$ \cite{MPbook}. For example for a system with four transitions ($n=3$) we see
\begin{equation*}
A_0^{\otimes2}= \left( \begin{array}{cccc}
 \epsilon &\epsilon &\epsilon &\epsilon \\
 \epsilon &\epsilon &\epsilon &\epsilon \\
\tau_1\otimes\tau_2&\epsilon &\epsilon &\epsilon \\
 \epsilon &\tau_2\otimes\tau_3&\epsilon &\epsilon 
\end{array} \right),~~
A_0^{\otimes3}= \left( \begin{array}{cccc}
 \epsilon &\epsilon &\epsilon &\epsilon \\
 \epsilon &\epsilon &\epsilon &\epsilon \\
 \epsilon &\epsilon &\epsilon &\epsilon \\
\tau_1\otimes\tau_2\otimes\tau_3 & \epsilon &\epsilon &\epsilon 
\end{array} \right);
\end{equation*}
i.e., the number of non-zero entries decreases as the power increases, and actually 
\begin{equation*}
A_0^*=\bigoplus_{k=0}^{\infty} A_0^{\otimes k}=\bigoplus_{k=0}^{n} A_0^{\otimes k}.
\end{equation*}
The Petri net describing translation can therefore be described by the equation
\begin{equation}
\mathbf{x}(k)=B\otimes \mathbf{x}(k-1),
\end{equation}
where $B=A_0^*\otimes A_1$ is given by 
\begin{align}
B= \left( \begin{array}{cccccc}
\tau_{\mathrm{in}} &0 &\epsilon &\epsilon&\cdots &\epsilon \\[0.8em]
\tau_{\mathrm{in}}\otimes\tau_1 &\tau_1 &0 &\epsilon&\cdots&\epsilon \\[0.8em]
\tau_{\mathrm{in}}\otimes\tau_1\otimes\tau_2  &\tau_1\otimes\tau_2 &\tau_2 &0&\ddots&\vdots \\[0.8em]
\tau_{\mathrm{in}}\otimes\tau_1\otimes\tau_2\otimes\tau_3  &\tau_1\otimes\tau_2\otimes\tau_3 & \tau_2\otimes\tau_3 &\ddots&\ddots&\epsilon \\[0.8em]
\vdots &\vdots&\vdots&\ddots&\tau_n&0\\[0.8em]
\tau_{\mathrm{in}}\otimes\tau_1\otimes\dots\otimes\tau_n &\tau_1\otimes\dots\otimes\tau_n&\tau_2\otimes\dots\otimes\tau_n&\cdots&\tau_{n-1}\otimes\tau_n&\tau_n\oplus\tau_{\mathrm{out}}
\end{array} \right). \label{B}
\end{align}

\begin{figure}
\centering
\includegraphics{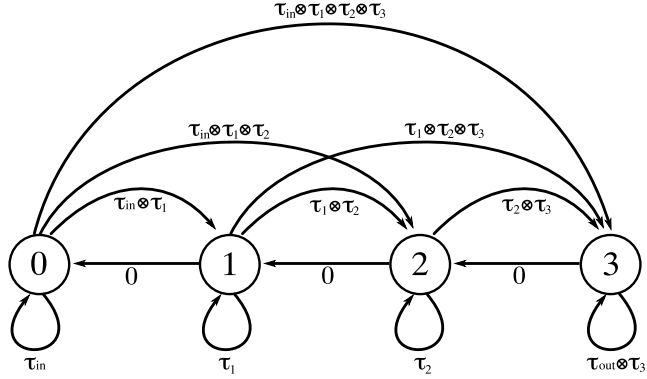}
\caption{Diagram showing the weighted, directed graph $\mathcal{G}(B)$ corresponding to the matrix $B$ (Eq. (\ref{B})) with $n=3$.  \label{graphB} }
\end{figure}

The eigenvalue and cyclicity of the matrix $B$ can be found from the corresponding graph $\mathcal{G}(B)$. In Fig. \ref{graphB} we show $\mathcal{G}(B)$ for the example of $n=3$. We note that the graph is strongly connected, so $B$ is irreducible and has exactly one eigenvalue (Theorem \ref{T:eigen}). Due to non-zero elements on the diagonal of $B$, there exist circuits of length one, so the cyclicity of the graph is $\sigma_{\mathcal{G}}=1$. By inspection of the graph it is clear that one of the circuits of length one will always have a mean circuit weight $\bar{w}_\gamma$ equal to the maximum mean circuit weight. That is to say, the maximum mean circuit weight, and therefore the eigenvalue of $B$ will be equal to whichever of the waiting times is largest, i.e., 
\begin{equation*}
\lambda=\max\{\tau_{in},\tau_{out},\tau_i|1\leq i\leq n\}.
\end{equation*}
It is also clear then that the critical graph of $B$ must always contain a circuit of length one, so $\sigma_B=1$.

Due to the cyclicity theorem (Theorem \ref{T:cyc}), we conclude that for large $k$ the vector $\mathbf{x}(k)$ is an eigenvector of $B$. Regardless of the initial condition $\mathbf{x}(0)$, the Petri net describing translation always reaches a steady state described by
\begin{equation}
\mathbf{x}(k)=\lambda \otimes \mathbf{x}(k-1), \label{sseig}
\end{equation}
for $k\geq K_0$, where $K_0$ is the transient time of $A$. The component of the eigenvector $x_i(k)$ gives the time at which the $i$th transition fires for the $k$th time. The $n$th component of this vector equation tells us that the transition corresponding to completion of an amino acid chain will fire at time intervals of $\lambda$; thus the eigenvalue of $A$ can be interpreted as the protein production time. As noted above, another way of interpreting the eigenvector is that $x_i(k)$ is the time at which the $k$th token (ribosome) leaves the $i$th place (codon); the token enters the $i$th place at time $x_{i-1}(k)$ and the maximum length of time which a ribosome can stay at a given codon is $\lambda$. The proportion of time for which there is a token occupying the $i$th place in the steady state, i.e. the occupation density, is therefore given by
\begin{equation}
\rho_i=\frac{1}{ \lambda}(x_i- x_{i-1} )  ~~~ i=1,\dots n, \label{rhodef}
\end{equation}
where $\{x_i|0\leq i\leq n\}$ are components of \emph{any} eigenvector of $B$, since $x_i(k)-x_{i-1}(k)=x_i(l)-x_{i-1}(l)$ for all $k,l\geq K_0$. We note that $\rho_i$ are the components of a vector of length $n$. 

It now remains to evaluate the eigenvector. If we define $\mathbf{u}=\mathbf{x}(k-1)$, then from Eq. (\ref{sseig}), $\mathbf{x}(k)=\lambda\otimes \mathbf{u}$. We can then use this to evaluate the eigenvalue using Eq. (\ref{recurs}) which when written as components gives
\begin{align}
\lambda \otimes u_0 &= \tau_{\mathrm{in}} \otimes u_0 \oplus u_1, \label{one}\\
\lambda \otimes u_i &= \lambda\otimes \tau_i \otimes u_{i-1} \oplus u_{i+1}, ~~~~~i=1,\dots n-1, \\
\lambda \otimes u_n &= \lambda\otimes\tau_n \otimes u_{n-1} \oplus \tau_{\mathrm{out}} \otimes u_n. \label{three}
\end{align}
Thus the eigenvectors takes a different form depending on the eigenvalue. We identify the following three cases:

\subsubsection{Case (i): $\lambda=\tau_{in}$}

We first consider the case where the parameters are such that the waiting time for initiation of translation is longer than that for tRNA capture and termination, i.e., $\tau_{\mathrm{in}}>\tau_{out},\tau_i$ for $1\leq i\leq n$. We shall denote this the \emph{entry limited regime}, as it represents the situation where entry of the ribosomes into the system is the rate limiting process. Replacing $\lambda\rightarrow\tau_{\mathrm{in}}$ in  Eqs. (\ref{one}-\ref{three}) gives
\begin{align}
 \tau_{\mathrm{in}}\otimes u_0 &= \tau_{\mathrm{in}} \otimes u_0 \oplus u_1, \label{i1}\\
 \tau_{\mathrm{in}}\otimes u_i &= \tau_{\mathrm{in}}\otimes \tau_i \otimes u_{i-1} \oplus u_{i+1}, ~~~~~i=1,\dots n-1, \label{i2}\\
 \tau_{\mathrm{in}}\otimes u_n &= \tau_{\mathrm{in}}\otimes\tau_n \otimes u_{n-1} \oplus \tau_{\mathrm{out}} \otimes u_n. \label{i3}
\end{align}
From (\ref{i3}), since $\tau_{\mathrm{in}} \neq \tau_{\mathrm{out}}$, for consistency it must be the case that $\tau_{\mathrm{in}} \otimes \tau_n\otimes u_{n-1} > \tau_{\mathrm{out}} \otimes u_n$, and therefore
\begin{equation}
u_n=\tau_n \otimes u_{n-1}. \label{case1_3}
\end{equation}
From (\ref{i2}), taking $i=n-1$ gives
\begin{equation*}
\tau_{\mathrm{in}} \otimes u_{n-1} = \tau_{\mathrm{in}} \otimes \tau_{n-1} \otimes u_{n-2} \oplus u_{n}.
\end{equation*}
Since $\tau_{\mathrm{in}}\neq \tau_n$, in order for this to be consistent with Eq. (\ref{case1_3}) it must be the case that $\tau_{\mathrm{in}} \otimes \tau_j \otimes u_{n-2} > u_{n}$, meaning
\begin{align*}
u_{n-2}&=\tau_{n-1}^{\otimes(-1)} \otimes u_{n-1} \nonumber\\
&=\tau_{n-1}^{\otimes(-1)} \otimes\tau_{n}^{\otimes(-1)} \otimes u_n.
\end{align*}
Continuing to use Eq. (\ref{i2}) in this way we can find expressions for each component of $\mathbf{u}$ in terms of the $n$th component. The eigenvector is given by
\begin{equation*}
\mathbf{u}=\left(
\begin{array}{c}
\tau_1^{\otimes(-1)}\otimes \dots \otimes \tau_n^{\otimes(-1)} \otimes u_n \\
\tau_2^{\otimes(-1)}\otimes \dots \otimes \tau_n^{\otimes(-1)} \otimes u_n \\
\vdots \\
\tau_n^{\otimes(-1)} \otimes u_n \\
u_n
\end{array} \right).
\end{equation*}
Using Eq. (\ref{rhodef}) we find $\rho_i=\tau_i/\tau_{\mathrm{in}}$.

\subsubsection{Case (ii): $\lambda=\tau_{out}$}

In the case where $\tau_{out}>\tau_{in},\tau_i$ for $1\leq i\leq n$, the time it takes a ribosome to leave the mRNA is the limiting process, i.e., this is the \emph{exit limited regime}. Following the same method as above we make the replacement $\lambda\rightarrow\tau_{\mathrm{out}}$ in Eqs. (\ref{three}) giving
\begin{align}
 \tau_{\mathrm{out}}\otimes u_0 &= \tau_{\mathrm{in}} \otimes u_0 \oplus u_1, \label{ii1}\\
 \tau_{\mathrm{out}}\otimes u_i &= \tau_{\mathrm{out}}\otimes \tau_i \otimes u_{i-1} \oplus u_{i+1}, ~~~~~i=1,\dots n-1, \label{ii2}\\
 \tau_{\mathrm{out}}\otimes u_n &= \tau_{\mathrm{out}}\otimes\tau_n \otimes u_{n-1} \oplus \tau_{\mathrm{out}} \otimes u_n. \label{ii3}
\end{align}
 This time we start with the first of the three equations, and note that since $\tau_{\mathrm{in}} \neq \tau_{\mathrm{out}}$, it must be the case that $u_1 > \tau_{\mathrm{in}}\otimes u_0$; therefore
\begin{equation}
u_1=\tau_{\mathrm{out}} \otimes u_0. \label{case2_1}
\end{equation}
We then consider Eq. (\ref{ii2}); for $i=1$ this gives
\begin{equation*}
\tau_{\mathrm{out}}\otimes u_1 = \tau_{\mathrm{out}}\otimes \tau_1 \otimes u_0 \oplus u_2.
\end{equation*}
In order for this to be consistent with Eq. (\ref{case2_1}), it must be the case that $u_2>\tau_{\mathrm{out}}\otimes\tau_1\otimes u_0$, meaning
\begin{align*}
u_2&=\tau_{\mathrm{out}} \otimes u_1 \nonumber\\
&=\tau_{\mathrm{out}}^{\otimes2} \otimes u_0.
\end{align*}
We continue using (\ref{ii2}) to find the other components of the vector; the result is
\begin{equation*}
\mathbf{u}=\left(
\begin{array}{c}
u_0\\
\tau_{\mathrm{out}} \otimes u_0 \\
\tau_{\mathrm{out}}^{\otimes 2}  u_0 \\
\vdots \\
\tau_{\mathrm{out}}^{\otimes n}  u_0
\end{array} \right),
\end{equation*}
leading to occupation densities $\rho_i=1$.

\subsubsection{Case (iii): $\lambda=\tau$}

The final case is the \emph{elongation limited regime}, where one or more of the internal waiting times has a value $\tau$, with $\tau>\tau_{\mathrm{in}},\tau_{\mathrm{out}},\{\tau_i\neq \tau \}$. We assume initially that there are two places with waiting times equal to $\tau$ deep in the bulk of the mRNA, denoting the first the $p$th, and the second the $q$th. That is to say $\tau_p,\tau_q=\tau$ where $1<p<q<n$. We again take Eqs. (\ref{three}) and this time make the replacement $\lambda\rightarrow\tau$, giving
\begin{align}
\tau \otimes u_0 &= \tau_{\mathrm{in}} \otimes u_0 \oplus u_1, \label{three1}\\
\tau \otimes u_i &= \tau\otimes \tau_i \otimes u_{i-1} \oplus u_{i+1}, ~~1\leq i < n,\label{three2} \\
\tau \otimes u_n &= \tau\otimes\tau_n \otimes u_{n-1} \oplus \tau_{\mathrm{out}} \otimes u_n. \label{three3}
\end{align}
We start with Eq. (\ref{three1}), which since $\tau_{\mathrm{in}}\neq\tau$, gives
\begin{equation*}
u_1=\tau \otimes u_0.
\end{equation*}
Moving onto Eq. (\ref{three2}) and taking $i=1$, if $\tau_1\neq\tau$ it must be the case that
\begin{align*}
u_2&=\tau \otimes u_1 \\
&=\tau^{\otimes 2}\otimes u_0.
\end{align*}
We can continue using Eq. (\ref{three2}) in this way until we reach the $p$th waiting time, since $\tau_p=\tau$. Taking $i=p-1$ and then $i=p$ in Eq. (\ref{three2}) we find respectively
\begin{align*}
\tau\otimes u_{p-1} &=u_p\\
\tau \otimes u_p& = \tau^{\otimes 2} u_{p-1} \oplus u_{p+1}.
\end{align*}
Since neither term on the right hand side of the second equation would contradict the first, all this can tell us is that 
\begin{equation}
u_{p+1}\leq \tau^{\otimes 2} u_{p-1} . \label{ineq1}
\end{equation} 
We now consider Eq. (\ref{three3}); since $\tau_{\mathrm{out}}\neq \tau$ it must be the case that
\begin{equation}
u_n=\tau_n \otimes u_{n-1}.\label{un}
\end{equation}
Using Eq. (\ref{three2}) with $i=n-1$ gives
\begin{equation*}
\tau\otimes u_{n-1} = \tau\otimes\tau_{n-1}\otimes u_{n-2} \oplus u_n.
\end{equation*}
If $\tau_n\neq \tau$ then in order not to contradict Eq. (\ref{un}), it must be the case that $\tau\otimes\tau_{n-1}\otimes u_{n-2} \geq u_n$. Therefore 
\begin{equation*}
u_{n-1}=\tau_{n-1}\otimes u_{n-2}.
\end{equation*}
We can continue to iterate backwards using Eq. (\ref{three2}) until we reach the $q$th waiting time $\tau_q=\tau$. Taking $i=q$ and then $i=q-1$ gives respectively
\begin{align*}
u_q &= \tau \otimes u_{q-1} \\
\tau\otimes u_{q-1} &= \tau\otimes \tau_{q-1} \otimes u_{q-2} \oplus u_q.
\end{align*}
Again the second equation can only give an inequality
\begin{equation}
u_q \geq \tau\otimes \tau_{q-1} \otimes u_{q-2}. \label{ineq2}
\end{equation}
We are therefore left with two inequalities, Eqs. (\ref{ineq1}) and (\ref{ineq2}) which cannot alone tell us about the $p$the and the $q$th components of the eigenvector. We find that if we assume that one of these inequalities is actually an equality, then we can continue using Eq. (\ref{three2}) to find all of the components of the eigenvector consistently with all of the above equations. This leads to two eigenvectors depending on which inequality we set equal
\begin{align*}
\mathbf{u}=\left(
\begin{array}{c}
u_0\\
\tau\otimes u_0 \\
\tau^{\otimes 2} \otimes u_0 \\
\vdots \\
\tau^{\otimes q}  \otimes u_0 \\
\tau_{q+1}\otimes\tau^{\otimes q} \otimes u_0 \\
\tau_{q+1}\otimes \tau_{q+2}\otimes\tau^{\otimes q} \otimes u_0 \\
\vdots \\
\tau_{q+1}\otimes\dots\otimes\tau_{n}\otimes\tau^{\otimes q} \otimes u_0
\end{array} \right),
~~\mathrm{or}~~
\mathbf{u}=\left(
\begin{array}{c}
u_0\\
\tau\otimes u_0 \\
\tau^{\otimes 2} \otimes u_0 \\
\vdots \\
\tau^{\otimes p} \otimes u_0 \\
\tau_{p+1}\otimes\tau^{\otimes p} \otimes u_0 \\
\tau_{p+1}\otimes \tau_{p+2}\otimes\tau^{\otimes p} \otimes u_0 \\
\vdots \\
\tau_{p+1}\otimes\dots\otimes\tau_{n}\otimes\tau^{\otimes p} \otimes u_0
\end{array} \right),
\end{align*}
where we recall that $p<q$ and the $p$th and $q$th are the upstream and downstream most codons with waiting times equal to $\tau$. Any max-plus linear combination either of these two vectors is an eigenvector of $B$. If we consider the critical graph of $B$, if there are $M$ non-adjacent codons with waiting times equal to $\tau$, then this graph will contain $M$ m.s.c.s., and therefore (via Theorem \ref{mscs}) there are $M$ linearly independent eigenvectors. We note that the critical graphs in cases (i) and (ii) have only one m.s.c.s..

Each of the two eigenvectors above gives rise to a different occupation density profile using Eq. (\ref{rhodef}), i.e., 
\begin{equation*}
\rho_i=\left\{ \begin{array}{ll}
1& \mathrm{for}~1\leq i \leq q, \\
\tau_i/\tau & \mathrm{for}~q<i\leq n,
\end{array} \right.
\end{equation*}
and 
\begin{equation*}
\rho_i=\left\{ \begin{array}{ll}
1& \mathrm{for}~1\leq i \leq p, \\
\tau_i/\tau & \mathrm{for}~p<i\leq n,
\end{array} \right.
\end{equation*}
which corresponds to queues of ribosomes behind the $q$th and the $p$th codons respectively. In mRNAs with more than two isolated codons with waiting times equal to $\tau$, there will be solutions with queues behind each of these ``slowest'' codons. We will show in the next section that only one of the solutions is realised, and we reserve further discussion until Sec. \ref{sec:het}.

\section{Simulation Results}\label{sec:results}

In this section we compare the results from the max-plus algebra detailed above with direct numerical simulation of the Petri net shown in Fig. \ref{mRNA_petri}(a). We conduct the simulations by considering an initial state where all bottom row places contain a token and all top row places are empty (the first and last places in the chain both contain one token). We then move forward in time from one event (firing of a transition) to the next, updating token positions at each step. We allow the system to reach a steady state, and then examine the protein production time $P$, and the codon occupation density of each place $\rho_i$, as well as the mean density, defined as
\begin{equation}
\rho=\frac{1}{n}\sum_{i=1}^{n} \rho_i.
\end{equation} 

We first consider a very simple mRNA where each codon is assumed to be identical, i.e., it codes for the same tRNA, and the waiting time for each is the same. The internal hopping rates are chosen to be $\tau_i=\bar{\tau}$ for $1\leq i\leq n$, and $\bar{\tau}$ is used as the unit of time.  We then measure $P$, $\rho_i$ and $\rho$ for different values of $\tau_{\mathrm{in}}$ and $\tau_{\mathrm{out}}$. 

In Sec. \ref{sec:het} we consider real mRNA sequences from the \emph{S. Cerevisiae} genome. The waiting time for a particular tRNA species is assumed to be inversely proportional to the concentration of those molecules found in a typical cell. We estimate this from the gene copy number of each tRNA \cite{Percudani1997}. We choose the internal hopping rates such that the average of the waiting times is equal to $\bar{\tau}$, and again use this as the unit of time.

\subsection{Homogeneous mRNAs}\label{sec:hom}

\begin{figure}
\centering
\includegraphics{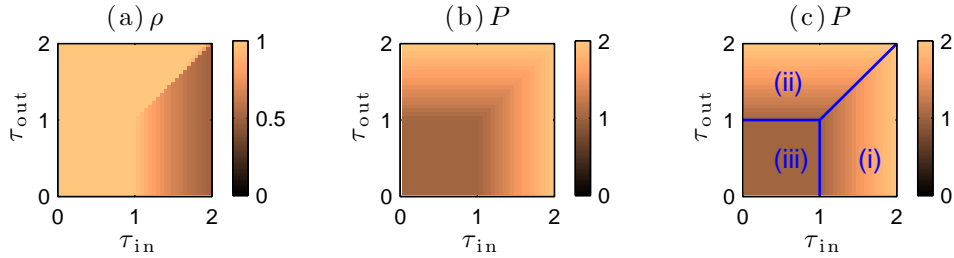}
\caption{Colour on-line. Colour maps showing how (a) the mean ribosome density $\rho$ and (b)
  protein production time $P$ vary as a function of $\tau_{\mathrm{in}}$
  and $\tau_{\mathrm{out}}$, for a uniform mRNA where all codons have
  the same tRNA capture time $\tau_i=\bar{\tau}$. All times are in units of
  $\bar{\tau}$. Plot (c) shows the results for $P$ again, but also includes in blue the boundaries between the three regimes as determined from the max-plus algebra. \label{hom_phase} }
\end{figure}

\begin{figure}
\centering
\includegraphics{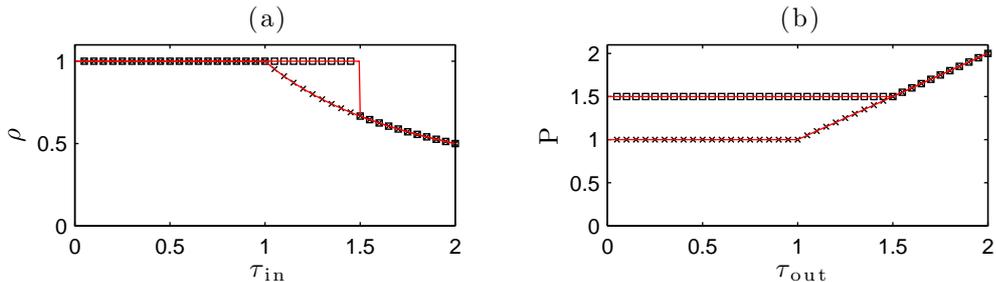}
\caption{Colour on-line. Comparison of numerical and analytic results for a uniform mRNA. Plots showing (a) the mean density $\rho$ and (b) the protein production time $P$ at different values of $\tau_{\mathrm{in}}$ and $\tau_{\mathrm{out}}$. Squares show numerical results for $\tau_{\mathrm{out}}=1.5\bar{\tau}$, and crosses those for $\tau_{\mathrm{out}}=0.5\bar{\tau}$ at different values of $\tau_{\mathrm{in}}$. Red lines show the analytic results from the max-plus algebra. All times are shown in units of $\bar{\tau}$. \label{hom2} }
\end{figure}

Figure \ref{hom_phase} shows colour maps for the protein
production time and mean density as a function of $\tau_{\mathrm{in}}$ and $\tau_{\mathrm{out}}$ for a uniform mRNA of length $n=500$, as generated from numerical simulations. There are three regimes, with the dynamics depending on the relative magnitudes of $\bar{\tau}$, $\tau_{\mathrm{in}}$, and $\tau_{\mathrm{out}}$. These correspond to the three cases found from the max-plus algebra, i.e., the initiation, termination and tRNA capture limited regimes. We summarise the results in  Table \ref{tab}.
Figure \ref{hom2} shows how $P$ and $\rho$ vary as a function of (a)
$\tau_{\mathrm{in}}$ for fixed $\tau_{\mathrm{out}}$, and (b)
$\tau_{\mathrm{out}}$ for fixed $\tau_{\mathrm{in}}$. The
numerical and analytic results match exactly.

\begin{table}
\centering
{\small
\begin{tabular}{|l |c|c|c|} \hline

& $\begin{array}{c} \mathrm{(i) ~Initiation ~ limited} \\ (\tau_{\mathrm{in}}>\tau_{\mathrm{out}},\bar{\tau}) \end{array} $
& $\begin{array}{c} \mathrm{ (ii) ~Termination ~ limited} \\ (\tau_{\mathrm{out}}>\tau_{\mathrm{in}},\bar{\tau}) \end{array} $
& $\begin{array}{c} \mathrm{ (iii) ~tRNA ~ capture ~ limited} \\ (\bar{\tau}>\tau_{\mathrm{in}},\tau_{\mathrm{out}}) \end{array}$\\\hline 

Protein production time $P$ & $\tau_{\mathrm{in}}$ & $\tau_{\mathrm{out}}$ & $\bar{\tau}$\\\hline
Occupation density $\rho_i$ & $\bar{\tau}/\tau_{\mathrm{in}}$ & 1 & 1 \\\hline
Mean density $\rho$ & $\bar{\tau}/\tau_{\mathrm{in}}$ & 1 & 1 \\\hline
\end{tabular}
}
\caption{Details of each regime as given by the
  max-plus algebra, for a uniform mRNA, where the tRNA capture waiting times are $\tau_i=\bar{\tau}$ for $1\leq i\leq n$. These are found to exactly match the numerical simulation results. \label{tab} }
\end{table}

\subsection{Heterogeneous mRNAs}\label{sec:het}

\begin{figure}
\centering
\includegraphics{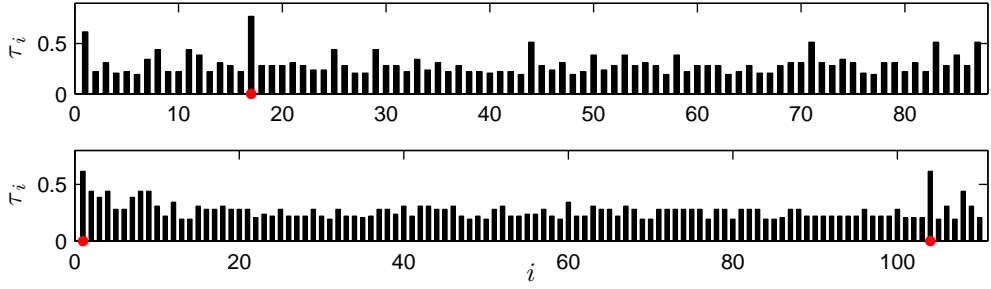}
\caption{Colour on-line. Waiting times for each codon on mRNAs A (top) and B (bottom). Red dots indicate codons which have the largest waiting time, which for mRNA A is $\tau=0.7696\bar{\tau}$ and for mRNA B is $\tau=0.6155\bar{\tau}$. Note that each waiting time $\tau_i$ has a value from the set $\{s_i|1\leq i \leq 41\}$. Whilst $(1/41)\sum_{i=1}^{41} s_i=\bar{\tau}$, in general $(1/n)\sum_{i=1}^{n} \tau_i \neq \bar{\tau}$.  \label{taui} }
\end{figure}

We now consider mRNA sequences from genes YJL136C and YDR382W in \emph{S. Cerevisiae}, which we here on denote mRNA A and mRNA B, respectively. For a given mRNA of length $n$ codons, each of the waiting times $\{\tau_i|1\leq i\leq n\}$ takes a value chosen from the set of 41 times $\{s_i|1\leq i\leq41\}$ corresponding to each of the 41 tRNA/codon species, and $(1/41)\sum_{i=1}^{41}s_i=\bar{\tau}$. We take $\bar{\tau}$ as the unit of time; biological experiments estimate that $\bar{\tau}\approx0.1~\mathrm{s}^{-1}$ \cite{Albertsbook}. Figure \ref{taui} shows the waiting times at each codon for each of the sequences.

\begin{figure}
\centering
\includegraphics{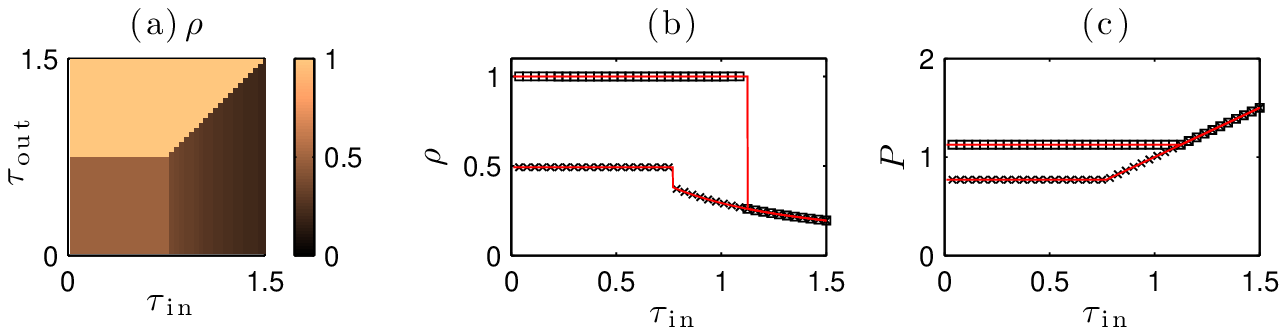}
\caption{Colour on-line. Comparison of numerical and analytic results for mRNA A showing values of $\rho$ and $P$ for different values of $\tau_{\mathrm{in}}$ and $\tau_{\mathrm{out}}$. In (a) the three regimes are clearly visible. In (b) and (c) squares show numerical results for $\tau_{\mathrm{out}}=1.125\bar{\tau}$ and crosses those for $\tau_{\mathrm{out}}=0.375\bar{\tau}$. Lines show the analytic results from the max-plus algebra. For this mRNA $\tau=0.7696\bar{\tau}$. \label{hetres1} }
\end{figure}

\begin{figure}
\centering
\includegraphics{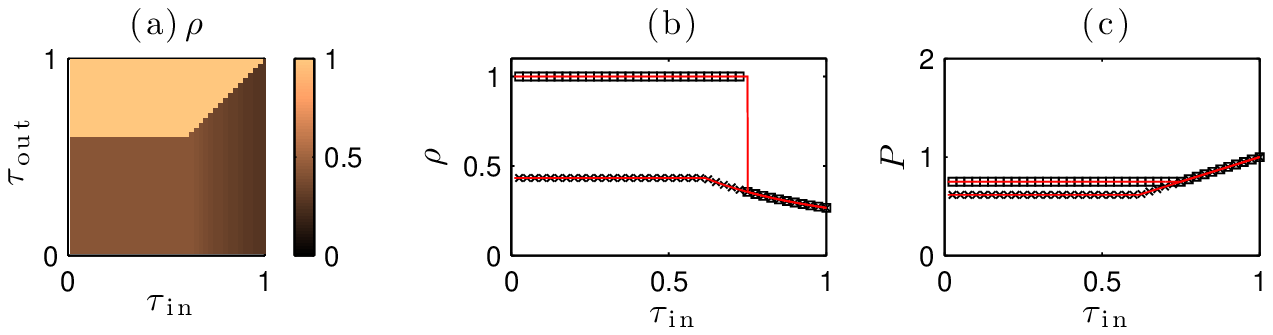}
\caption{Colour on-line. Comparison of numerical and analytic results for mRNA B showing values of $\rho$ and $P$ for different values of $\tau_{\mathrm{in}}$ and $\tau_{\mathrm{out}}$. In (a) the three regimes are clearly visible. In (b) and (c) squares show numerical results for $\tau_{\mathrm{out}}=0.75\bar{\tau}$ and crosses those for $\tau_{\mathrm{out}}=0.25\bar{\tau}$. Lines show the analytic results from the max-plus algebra. For this mRNA $\tau=0.6155\bar{\tau}$.\label{hetres2} }
\end{figure}

Figures \ref{hetres1} and \ref{hetres2} show simulation results for $P$ and $\rho$ for each mRNA respectively. We again observe three regimes depending on the relative magnitude of $\tau_{\mathrm{in}}$, $\tau_{\mathrm{out}}$, and the largest of the $\tau_i$ which we denote $\tau$. Again the simulation results exactly match those from the max-plus algebra, which are summarised in Table \ref{nonu_tab}.

\begin{table}
\centering
{\small
\begin{tabular}{|l |c|c|c|} \hline

& $\begin{array}{c} \mathrm{(i) ~Initiation ~ limited} \\ (\tau_{\mathrm{in}}>\tau_{\mathrm{out}},\tau) \end{array} $
& $\begin{array}{c} \mathrm{ (ii) ~Termination ~ limited} \\ (\tau_{\mathrm{out}}>\tau_{\mathrm{in}},\tau) \end{array} $
& $\begin{array}{c} \mathrm{ (iii) ~tRNA ~ capture ~ limited} \\ (\tau>\tau_{\mathrm{in}},\tau_{\mathrm{out}}) \end{array}$\\\hline 

Protein production time $P$ & $\tau_{\mathrm{in}}$ & $\tau_{\mathrm{out}}$ & $\tau$\\\hline
Occupation density $\rho_i$ & $\tau_i/\tau_{\mathrm{in}}$  & 1 & 
$\begin{array}{rc} 1 &\mathrm{for}~ 1\leq i\leq p,\\ \tau_i/\tau &\mathrm{for}~ p<i\leq n \end{array}$ \\\hline
{\small Mean density} $\rho$ & $\displaystyle \frac{1}{n \tau_{\mathrm{in}}} \sum_{i=1}^{n} \tau_i $ & 1 & $\displaystyle \frac{p}{n} + \frac{1}{n \tau} \sum_{i=p+1}^{n} \tau_i$\\\hline
\end{tabular}
}
\caption{Details of each phase as given by the max-plus algebra, for a non-uniform mRNA. Here $\tau$ is the longest internal waiting time ($\tau=\max\{\tau_i|1\leq i\leq n\}$), and $p$ is the position of the leftmost codon with waiting time equal to $\tau$. \label{nonu_tab} }
\end{table}

\begin{figure}
\centering
\includegraphics{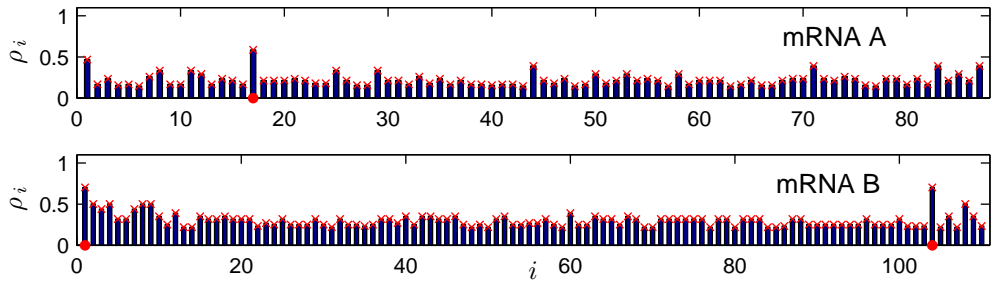}
\caption{Colour on-line. Occupation density profiles for each mRNA when parameters are chosen such that the system is in the entry limited phase. Parameters for mRNA A are $\tau_{\mathrm{in}}=1.3125\bar{\tau},~\tau_{\mathrm{out}}=0.0375\bar{\tau},~\tau=0.7969\bar{\tau}$, and for mRNA B $\tau_{\mathrm{in}}=0.875\bar{\tau},~\tau_{\mathrm{out}}=0.025\bar{\tau},~\tau=0.6155\bar{\tau}$. Bars show numerical results and crosses show results from the max-plus algebra. Red dots indicate the locations of the slowest codons.\label{hetrho1} }
\end{figure}

\begin{figure}
\centering
\includegraphics{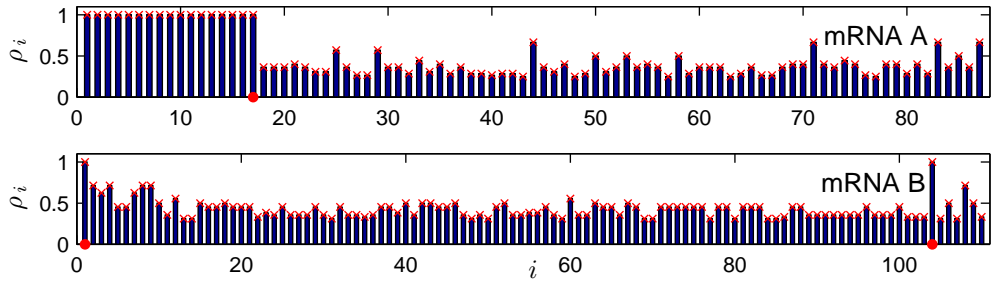}
\caption{Colour on-line. Occupation density profiles for each mRNA when parameters are chosen such that the system is in the tRNA capture limited phase. Parameters for mRNA A are $\tau_{\mathrm{in}}=0.0375\bar{\tau},~\tau_{\mathrm{out}}=0.0375\bar{\tau},~\tau=0.7969\bar{\tau}$, and for mRNA B $\tau_{\mathrm{in}}=0.025\bar{\tau},~\tau_{\mathrm{out}}=0.025\bar{\tau},~\tau=0.6155\bar{\tau}$. Bars show numerical results and crosses show results from the max-plus algebra. Red dots indicate the locations of the slowest codons. \label{hetrho2} }
\end{figure}

In two of the three regimes the occupation density differs from codon to codon. In Fig. \ref{hetrho1} we show the density profile for each mRNA when parameters are chosen such that the system is in the entry limited phase. We note that for most codons $i$, $\rho_i<0.5$; in other models (i.e., the TASEP which we discuss in Sec. \ref{sec:tasep}) the equivalent regime is usually called the low density phase. In Fig. \ref{hetrho2} we show the density profiles for the tRNA capture limited phase. For codons located upstream of the most upstream codon with the longest waiting time ($\tau_p=\tau$) the density $\rho_i\approx 1$ for $i<p$; i.e., the leftmost instance of the slowest codon causes queueing of ribosomes.  In the case of mRNA B the first ``slowest codon'' is at $i=1$, so no queue is observed. Although from the max-plus algebra there are other possible solutions, only one is realised in the simulations. Instances of the slowest codon appearing further downstream do not give rise to queues, since the rate at which ribosomes arrive at those codons is too low. In terms of the max plus algebra, the different possible eigenvectors are reachable by different initial conditions \cite{butkovicbook}. The initial condition where are the codons are vacant corresponds to the initial vector $\mathbf{x}(0)$ where $x_0(0)=0$ and $x_i(0)=\epsilon$ for $i\neq0$; i.e. we start counting the elapsed time from $0$ at the first initiation event. Only the eigenvectors corresponding to a queue behind the most upstream slowest codon are reachable from this initial condition.

The other solution for the density profile of mRNA B can be accessed by making a change to the waiting times during the simulation. If we first allow the system to reach steady state, and then increase the waiting time on the slow codon at position $i=104$ by some small amount $\delta \tau$, then a queue will begin to form behind this codon. If we reverse the change after the system reaches steady state, then the queue behind codon $104$ will persist (data not shown).

\section{Introduction of stochasticity}\label{sec:stoch}

In the above work we have examined a deterministic version of the model, where rather than choosing waiting times from a Poisson distribution, we instead use the mean of the distribution. Simulation of a Petri net with waiting times chosen from a distribution is straightforward. The description using max-plus algebra is less so. At its essence the problem consists of a product of irreducible max-plus matrices such that $\mathbf{x}(k+1)=A(k)\otimes\mathbf{x}(k)$ for $k\geq0$, where $\{A(k)|k \in \mathbb{N}\}$ is an independent and identically distributed (i.d.d.) sequence of random matrices. Some work on such sequences can be found in the literature \cite{MPbook,Mairesse1997}, and it can be shown that for an i.d.d. sequence of matrices which have fixed support and are irreducible, a max-plus Lyapunov exponent exists, and this is equal to the asymptotic growth rate. However, there is currently no method for calculating this quantity, which in the present model would be related to the mean protein production time. For this reason, here we only briefly discuss how simulation results for a stochastic Petri net compare with the deterministic case.

\begin{figure}
\centering
\includegraphics{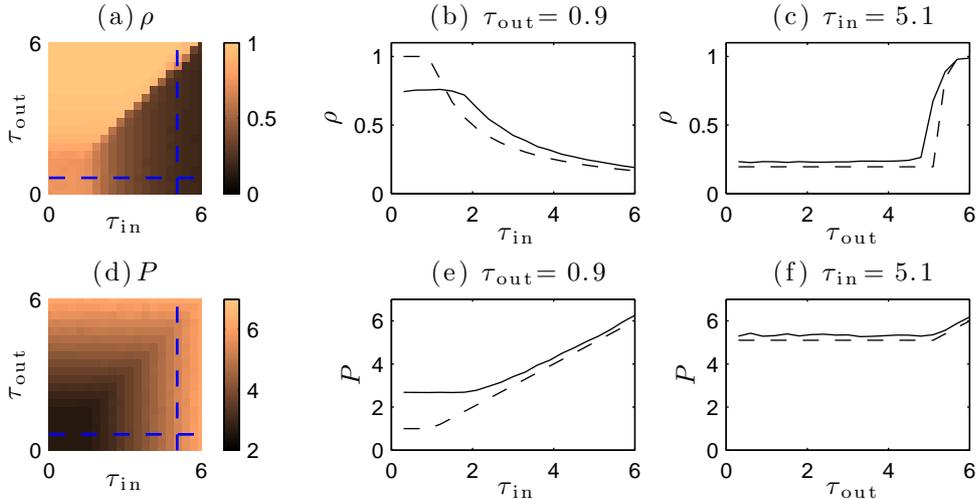}
\caption{Colour on-line. Simulation results for a stochastic Petri net representing a uniform mRNA of length $n=500$ codons, where each codon corresponds to the same tRNA species. Plots (a) and (d) show colour plots of the mean occupation density and protein production time interval respectively for different values of $\tau_{\mathrm{in}}$ and $\tau_{\mathrm{out}}$. The other plots show line graphs at cross sections through the colour maps as indicated by the dashed blue lines. Dashed lines in (b), (c), (e) and (f) show results for the corresponding deterministic Petri net. All times are given in units of the mean tRNA capture time $\bar{\tau}$. \label{stochfig} }
\end{figure}

We consider here a uniform mRNA of length $n=500$ codons, where waiting times for initiation, termination and tRNA capture are chosen from exponential distributions with means $\tau_{\mathrm{in}}$, $\tau_{\mathrm{out}}$ and $\bar{\tau}$ respectively. We perform simulations where we allow the system to reach steady state before measuring the time averaged protein production time and codon occupation density, which we again denote $P$ and $\rho$ respectively. As shown in Figs. \ref{stochfig}(a) and (d) we again see three regimes depending on which waiting time is the largest ($\tau_{\mathrm{in}}$ and $\tau_{\mathrm{out}}$ compared to $\bar{\tau}$). We note by comparing with Fig. \ref{hom_phase} that in the stochastic case the boundary for the tRNA capture limited regime is at $\tau_{\mathrm{in}},\tau_{\mathrm{out}}\approx2\bar{\tau}$, compared to $\tau_{\mathrm{in}},\tau_{\mathrm{out}}=\bar{\tau}$ in the deterministic case. Also in that regime, the density is significantly reduced compared to the deterministic case ($\rho\approx0.75$ compared to $\rho=1$) and the protein production time is increased ($P\approx2.7\bar{\tau}$ compared to $P=\bar{\tau}$). This is as expected since allowing the internal waiting times to vary will lead to gaps between ribosomes. The fact that the difference between the stochastic and deterministic models in this regime is so large can be attributed to the fact that the rate limiting process involves choosing waiting times for each of the $n=500$ places representing the codons. In the other regimes the effect of the stochasticity is less severe, as the limiting process involves only one place. Deep within the initiation or termination limited regimes $\rho$ and $P$ are almost the same as in the deterministic model. The general effect of stochastic waiting times in these regimes is that both density and protein production time increase. If we consider ribosomes occupying two consecutive codons, and if the waiting time drawn for the leading ribosome is long, it could hold up the ribosome behind it; if the time drawn for the leading ribosome is short it is unlikely to effect the ribosome behind it. Thus we expect ribosomes will in general move more slowly along the mRNA; as we shall see below, often the maximum of two exponentially distributed random numbers is what determines the overall behaviour.

For non-uniform mRNAs the effect of stochasticity in the initiation and termination limited regimes are similar: both $\rho$ and $P$ are increased compared to the deterministic model. In the tRNA capture regime the situation is more complicated, and features such as multiple queues are observed. Discussion of such phenomena is beyond the scope of the present paper. Here we only briefly consider one consequence of stochasticity which has played an important role in other translation models \cite{Chou2004,Dong2007}. It has been observed in a widely used alternative model for translation (the TASEP, as discussed in section \ref{sec:tasep}) that several slow codons in close proximity have a more dramatic effect on the protein production rate from an mRNA than slow codons in isolation. In the deterministic Petri net the token release time interval from a place corresponding to a slow codon (a slow place) which is in isolation is the same as that from the second of a pair of adjacent identical slow places. Once steady state has been reached the ``timer'' on the token in the first slow place will end at the same time as on the second, so a token will always enter the second place as soon as the previous token leaves - the time interval between tokens leaving the second place is exactly the waiting time for that place. The deterministic Petri net does not predict any effects due to clusters of slow codons. For a simple toy mRNA sequence it is possible to quantitatively predict how a cluster affects the protein production rate in the stochastic version of the Petri net model.

We consider an mRNA with very fast initiation and termination containing only very fast codons, except for a single slow codon somewhere in the bulk. We focus on the release time interval of tokens from the place corresponding to the slow codon. If the slow place has mean waiting time $\tau_{\mathrm{slow}}$, and we assume that all of the other waiting times are so fast that this place is refilled with a new token  practically immediately after the previous one leaves, then the mean release time interval for tokens from this place is equal to $\tau_{\mathrm{slow}}$.  
\begin{figure}
\centering
\includegraphics{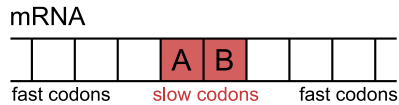}
\caption{Colour on-line. Schematic diagram showing a cluster of two slow codons labelled A and B. Ribosomes translate from left to right. All other codons have much shorter waiting times compared to A and B. \label{slow_cluster} }
\end{figure}
Now consider the same system, but with a pair of slow codons labelled A and B - see Fig. \ref{slow_cluster}. In the Petri net we denote the places corresponding to the slow codons place A and B. We denote the waiting times drawn from the (identical) distributions for each place $\tau^A_1$, $\tau^A_2,\dots$,  and $\tau^B_1$, $\tau^B_2,\dots$. We set $t=0$ when the system is in the steady state and a token enters place B (since the other codons are very fast, a token also enters place A at this time). The first token leaves place B at time $t=\tau^B_1$; the next token leaves B at $t=\tau^B_2+\max\{\tau^A_1,\tau^B_1\}$, and the next at $t=\tau^B_3+\max\{\tau^A_2,\tau^B_2\}+\max\{\tau^A_1,\tau^B_1\}$. That is, the $i$th token leaves place B at 
\begin{equation*}
t=\tau^B_i+\sum_{j=1}^{i-1} \max\{\tau^A_j,\tau^B_j\}.
\end{equation*}
The $i$th interval $\Delta_i$ between two consecutive tokens leaving B is therefore
\begin{equation*}
\Delta_i=\tau^B_i - \tau^B_{i-1} + \max\{\tau^A_{i-1},\tau^B_{i-1}\}.\label{Dt}
\end{equation*}
Taking the average, and noting that the distribution for each slow place is the same, gives
\begin{equation*}
\langle \Delta_i \rangle=\langle \max\{\tau^A_{i-1},\tau^B_{i-1}\} \rangle,
\end{equation*}
i.e., the mean release time interval is the mean of the maximum of two numbers drawn consecutively from the same exponential distribution. This can easily be shown\footnote{If two times $\tau_1$ and $\tau_2$ are random numbers where the probability that $\tau_i$ has a value $x$ is given by a Poisson distribution with mean $\tau_{\mathrm{slow}}$, i.e., , $P(\tau_i=x)=\tau_{\mathrm{slow}}^{-1}e^{-x/\tau_{\mathrm{slow}}}$, then the probability that $m=\max\{\tau_1,\tau_2\}$ has a value $x$ is given by $P(m=x)=P(\tau_1=x)\cdot P(\tau_2<x) + P(\tau_2=x)\cdot P(\tau_1<x)$. The mean value of $m$ is then just the expectation value of $P(m=x)$.} to be $3 \tau_{\mathrm{slow}}/2$, where $\tau_{\mathrm{slow}}$ is the mean waiting time of the slow codons. This agrees with our earlier assertion that, in general, stochastic waiting times leads to slower movement of ribosomes along the mRNA.

\section{Relationship to other translation models: TASEP}\label{sec:tasep}

An alternative framework for describing mRNA translation, is the totally asymmetric simple exclusion process, or TASEP \cite{MacDonald1968,Derrida1992,Chou2004}. In that model ribosomes are represented by particles which hop in one direction along a 1D lattice of sites, which represent the codons of the mRNA. The current model shows many similarities to the TASEP, but the dynamics unfold in an essentially different way. 

The dynamics of the TASEP are most often simulated via Monte Carlo methods using a random-sequential update rule, or using continuous time Monte Carlo \cite{Bortz1975}. In the former case, time is discretised and in each time step lattice sites are chosen at random with uniform probability. If the chosen site contains a particle and the next site is vacant the particle is advanced with some probability $p$; particles are injected from the leftmost site and removed from the rightmost site with probabilities $\alpha$ and $\beta$, respectively. For a lattice with $n$ sites this is repeated $n$ times in each time step. In the continuous time version, exactly one particle is moved in each time step but the particle is chosen based on its probability of hopping, and the duration of the time step drawn from the corresponding Poisson distribution. Both methods are equivalent in that they are the realisation of the usual master equation in continuous time; this last point means that multiplying all probabilities by a common factor leads only to a rescaling of time. Under some conditions the random sequential TASEP can be solved exactly \cite{Derrida1992}, and also a mean field treatment which ignore spatial correlations is often used.

The major difference between this and our timed Petri net picture is that in the TASEP only one particle can move at a certain instant in time. In the Petri net, as soon as any token move becomes possible, it is executed; i.e., events can happen simultaneously. Since in translation ribosomes actually take a finite time to move from one codon to the next, i.e., they are able to move at the same time, we propose that the Petri net more closely models the microscopic dynamics of this system. Results from the Petri net model can most easily be compared with the TASEP by identifying $\alpha\delta t=1/\tau_{\mathrm{in}}$, $\beta\delta t=1/\tau_{\mathrm{out}}$ and, in the case of a uniform mRNA, $p\delta t=1/\bar{\tau}$, where $\delta t$ is the time step. Although the uniform TASEP produces a very similar phase diagram to that of Fig \ref{hom_phase}, the subtle differences in the dynamics do lead to some important differences in the macroscopic behaviour. The analytic results from the max-plus treatment for the Petri net model and from a mean field approximation for the TASEP are summarised in \ref{TASEPtab}. As well as the density and protein production time differing in the two models, the boundaries of the regimes are also different. Specifically the boundaries for the tRNA capture limited regime (usually called the maximal current phase in the TASEP literature) are at $\tau_{\mathrm{in}} ,\tau_{\mathrm{out}} =2\bar{\tau}$ instead of $\bar{\tau}$ in the deterministic Petri net. We note that the phase boundaries in the TASEP are the same as those in the stochastic Petri net model discussed in the previous section; we could therefore ascribe this to the stochasticity. However in the tRNA capture limited regime we have $\rho=1/2$ and $P=4\bar{\tau}$ in the TASEP, compared to $\rho=1$ and $P=\bar{\tau}$ in the deterministic Petri net and $\rho\approx3/4$ and $P\approx2.7\bar{\tau}$ in the stochastic Petri net. The dynamics of the TASEP are still different to those of the stochastic Petri net.

\begin{table}
\centering
{\small
\begin{tabular}{|l |c|c|c| c|c|c|} \hline
\multirow{2}{*}{}&\multicolumn{3}{|c|}{Deterministic Petri Net} &\multicolumn{3}{|c|}{Random-sequential TASEP} \\ \cline{2-7}
&& $\rho$ & $P$ && $\rho$ & $P$ \\ \hline
(i) Initiation Limited &$\tau_{\mathrm{in}}>\tau_{\mathrm{out}},\bar{\tau}$&$\bar{\tau}/\tau_{\mathrm{in}}$&$\tau_{\mathrm{in}}$&$\begin{array}{l}\tau_{\mathrm{in}}>\tau_{\mathrm{out}} \\ \tau_{\mathrm{in}}>2\bar{\tau}\end{array}$&$\bar{\tau}/\tau_{\mathrm{in}}$& $\displaystyle\frac{\tau_{\mathrm{in}}^2}{\tau_{\mathrm{in}}-\bar{\tau}}$\\ \hline

(ii) Termination Limited &$\tau_{\mathrm{out}}>\tau_{\mathrm{in}},\bar{\tau}$&1&$\tau_{\mathrm{out}}$&$\begin{array}{l} \tau_{\mathrm{out}}>\tau_{\mathrm{in}} \\  \tau_{\mathrm{out}}>2\bar{\tau}\end{array} $ &$\displaystyle \bar{\tau}\left(1-\frac{1}{\tau_{\mathrm{out}}}\right)$& $\displaystyle\frac{\tau_{\mathrm{out}}^2}{\tau_{\mathrm{out}}-\bar{\tau}}$ \\ \hline

(iii) tRNA Capture Limited &$\tau_{\mathrm{out}},\tau_{\mathrm{in}}<\bar{\tau}$&1&$\bar{\tau}$&$\tau_{\mathrm{in}},\tau_{\mathrm{out}}<2\bar{\tau}$&$4\bar{\tau}$&  $1/2$ \\ \hline

\end{tabular}
}
\caption{Comparison of the results from the Petri net and TASEP models for a uniform mRNA. The TASEP results are from a mean field treatment which has been shown to very closely match random-sequential Monte Carlo simulations for large systems \cite{Derrida1992}. \label{TASEPtab} }
\end{table}

A TASEP with an ordered-sequential update has been studied by some authors \cite{Rajewsky1997,Rajewsky1998,Romano2009}, and is particularly applicable to, for example, traffic flow models, movement of molecular motors, and -- as we study here -- mRNA translation. This update rule corresponds more closely to the Petri net model. Sites are taken in turn from right to left and updated according to the hopping probabilities. However, this has remained a less favourable update rule since analytic approaches run into several difficulties, particularly in the physical interpretation of hopping probabilities. Unlike the random-sequential case, for the ordered update time is not continuous, i.e., scaling the probabilities does not lead to a simple scaling of time (for further discussion see \cite{Rajewsky1998}). The advantage of the Petri net picture is therefore clear: we can include the fact that multiple events can happen simultaneously, whilst retaining (at least in the deterministic case) an analytically soluble mathematical framework, i.e., algebra on the max-plus semi-ring.

Another difference between the current model and the TASEP is in the biological interpretation of the waiting time of the ribosome on each codon. In the random-sequential TASEP the hopping probability from a site is chosen based on the abundance of tRNAs which corresponds to that codon. A recent study \cite{Ciandrini2010} shows that this is in fact the wrong interpretation of the hopping probability. Ciandrini et al. have developed an extension to the TASEP where ribosomes take two internal states. They identify two times: the waiting time for a ribosome to capture the correct tRNA $\tau_{\mathrm{capture}}$, and the time it takes for the ribosome to physically move from one site to the next $\tau_{\mathrm{move}}$. The former of these depends on the availability of tRNAs whilst the latter does not, and they argue that the capture of the tRNA can occur independently of whether or not there is a vacancy to the right of the ribosome. A commonly held misconception in applying the TASEP to translation is that the model represents the limit $\tau_{\mathrm{move}}\rightarrow0$; Ciandrini's model shows that it is the opposite $\tau_{\mathrm{capture}}\rightarrow0$ limit which recovers the original TASEP model; importantly, this limit does not describe the biologically relevant regime ($\tau_{\mathrm{move}}\ll \tau_{\mathrm{capture}}$). In the timed Petri net the place waiting times correspond to $\tau_{\mathrm{capture}}$, hence we explicitly operate in the biologically relevant $\tau_{\mathrm{move}}\rightarrow0$ limit.

\section{Concluding Remarks}

We have presented here a new model of ribosome dynamics during mRNA translation. Algebra on the max-plus semi ring lends its self to describing systems in which discrete events occur depending on the fulfilment of conditions, and has previously be used to study, for example, distributed software systems, automated manufacturing or industrial control systems. We have shown here that biological systems represent another area where such methods can be applied. The Petri net is a useful tool for visually representing discrete event systems, and the wealth of previous work on describing Petri nets using max-plus has allowed us to quickly develop a framework for predicting the dynamical behaviour of elongating ribosomes given only the codon sequence of an mRNA.

The analytic treatment we have presented is applicable to a deterministic version of the model, and we have discussed the modifications to the system when ribosome waiting times are chosen from a distribution. The present work therefore represents the first step in a new direction for protein synthesis modelling. Whilst numerical simulation of a stochastic Petri net is straightforward, there are currently few methods or algorithms available for the study of sequences of i.d.d. max-plus matrices. Therefore this work also represents a new source of motivation in this area.

There is also scope for introducing features which bring the model closer to the biology. For example including a finite time for the physical movement of the ribosomes \cite{Ciandrini2010}, taking into account the fact that ribosomes cover more than one codon at a time \cite{Shaw2003,Shaw2004}, or allowing for a finite pool of ribosomes or other resources \cite{adams2008,Brackley2010,Brackley2010b}.

In summary we have presented a new model of translation which uses max-plus algebra to solve a Petri net description of the system. The microscopic dynamics are more realistic than in other translation models, and calculation of quantities such as density profiles and protein production rates is straightforward. Unlike other translation models the max-plus algebra also allows for exact analytic solutions for inhomogeneous mRNA sequences.

\bibliographystyle{unsrt}
\bibliography{petri.bib}

\end{document}